# Android Anti-forensics: Modifying CyanogenMod

Karl-Johan Karlsson
University of Glasgow
creideiki@lysator.liu.se

William Bradley Glisson
University of South Alabama
bglisson@southalabama.edu

**Abstract**

*Mobile devices implementing Android operating systems inherently create opportunities to present environments that are conducive to anti-forensic activities. Previous mobile forensics research focused on applications and data hiding anti-forensics solutions. In this work, a set of modifications were developed and implemented on a CyanogenMod community distribution of the Android operating system. The execution of these solutions successfully prevented data extractions, blocked the installation of forensic tools, created extraction delays and presented false data to industry accepted forensic analysis tools without impacting normal use of the device. The research contribution is an initial empirical analysis of the viability of operating system modifications in an anti-forensics context along with providing the foundation for future research.*

## 1. Introduction

The increasing integration of mobile smartphones, in today's digitally dependant, highly networked, communication based societies creates an environment that is conducive to encouraging anti-forensics activities. According to the International Telecommunications Union [23], at the end of 2011 there were almost six billion mobile phone subscriptions for a world population of seven billion. In the fourth quarter of 2012, 207.7 million smartphones were sold with Android capturing over 50% of the operating system market [13]. Smartphones can be described as general-purpose computers with an attached phone. As such, many people use smartphones for their daily consumption, storage and communications tasks. This makes smartphones a great source of forensic evidence while, simultaneously, presenting interesting analysis challenges.

Mobile smartphones are highly integrated devices that are built from non-standard components, running software which is often proprietary, undocumented and frequently changed. To perform a component-by-component analysis, an analyst would start by disassembling the phone and removing the surface mounted memory chips, which is a delicate and highly risky procedure. The memory chips can be read by standardized readers, but the interpretation of the data depends on the software running on the phone. A much easier method is to let the phone run, and access the data through the normal interfaces provided by the software. However, this presents a high risk of data being modified, both as a normal function of the phone and/or by specialized anti-forensic applications. The savings in time and effort gained by the utilization of normal interfaces are substantial enough that this technique is endorsed by the Association of Chief Police Officers (ACPO) [32] and the American National Institute of Standards and Technology [24].

Due to this acceptance, forensic analysts rely heavily on the correct functioning of the phone's software when performing analyses. Hence, altering functionality is a way of thwarting an analysis. Smartphones running operating systems such as Android and iOS are designed to allow the installation of third-party applications. This has allowed for the development of applications with anti-forensic functionality [7, 12, 27]. However, these applications have to work under the restrictions imposed by the operating system, such as application isolation and responsiveness demands. If anti-forensic modifications were to be made on a lower level, these restrictions would not apply in the same way, possibly making more advanced methods available. This idea promoted research into the hypothesis that *it is possible to modify the Android operating system to present false information to the forensics tools*. Several subsidiary research questions were identified in order to explore the hypothesis:

1. Which components of the Android operating system do the forensics tools trust?
2. Is it possible to modify these components to present false information?
3. Can the presence of a forensic analysis tool be detected?

4. Is it possible to make the presentation of false information reversible, such that the phone will revert to presenting the real information after the forensic analysis?

The research contribution is an initial empirical analysis of the viability of operating system modifications in an anti-forensics context along with providing the foundation for future research in this area. The paper is structured as follows: Section 2 discusses relevant approaches to smartphone anti-forensics. Section 3 presents the methodology, and the experimental design. Section 4 discusses the implementation and results. Section 5 draws conclusions from the research conducted and Section 6 presents future work.

## 2. Relevant Work

Recent research investigates the risk that mobile phones present to individual members of society [16] and to the business world [14]. It has also examined some of the challenges these devices present to forensic investigations [9, 17, 18, 25]. Hence, it is only a matter of time before individuals, organizations and businesses implement solutions to mitigate these risks through anti-forensics activities.

Anti-forensics in this case is broadly defined as "any attempts to compromise the availability or usefulness of evidence to the forensics process. Compromising evidence availability includes any attempts to prevent evidence from existing, hiding existing evidence or otherwise manipulating evidence to ensure that it is no longer within reach of the investigator" [19]. While several data extraction options exist for mobile devices, research has highlighted the fact that not all extraction solutions are equal nor do they necessarily provide the ability to validate results [15]. This can be attributed to an array of factors that include numerous mobile phone hardware configurations and vast numbers of devices in the market. Hence, the extractions that are most likely to be implemented with higher degrees of success are logical and manual extractions.

Android is a young operating system, with the first commercial device, the HTC Dream, also known as the T-Mobile G1, launched in September 2008 [31]. Hence, it is expected that the Android forensics and anti-forensics literature will not be as established as the ones for Windows PCs. Harris [19] classifies anti-forensics into four groups: hiding, destruction, source elimination and counterfeiting. Kessler [26] also categorizes anti-forensics into similar groups which consist of data hiding, artifact wiping, trail obfuscation, and attacks against processes and tools.

### 2.1 Data hiding

Data hiding on mobile devices will implement substantially similar approaches to that of personal computers like steganography, deleted files, and storing data in the cloud or in other users' storage space. The caveat with this approach on mobile devices is that recovery of deleted files depends on the file system. Many mobile devices use a version of Yet Another Flash File System (YAFFS) [1], which may be unsupported in commercial forensics tools.

Specific for Android is the separation between different applications enforced by the operating system. Every application is run as its own Linux user. Standard Linux file system permissions are used to ensure that no other application can read its files. This also applies to the applications uploaded to the phone by forensic analysis tools. This protection can only be bypassed if the phone is first rooted. If that is done, software can use the elevated privilege to read the entire file system.

On a non-rooted phone, then, information can be hidden by having an application store it somewhere secluded and restore it at a later time (such as when the user enters a password). This approach was tested by Distefano et al. [12]. Their program takes data from a number of standard databases on the phone (e.g. contact list, call logs, and SMS messages) and user-specified files, copies this data to files in the program's directory and deletes the originals. This approach also allows for quick mass deletion, since the Android package manager deletes all files private to an application if it is uninstalled.

They attempted to use the forensics tool Paraben Device Seizure [30], but found that this was incompatible with the phone they were using. Instead, they used backup programs, which require the phone to be rooted and perform a logical acquisition of the phone memory. As expected, these programs were able to read the private directory where the data had been stored. Had the phone not been rooted, the backup programs would not have worked [20].

Distefano, et al. [12] say nothing about how their data hiding program is triggered, nor when data is put back. They do, however, include test results for how long it takes for the hiding process to run. This was on the order of 10 seconds, depending on the amount of data to be hidden. This suggests that the hiding process is action sensitive, which would be the case if it was triggered by the connection or starting of a forensics tool. Presumably, the data would then be manually restored by the user after regaining control over the phone.

## 2.2 Artifact wiping

Artifact wiping is the act of overwriting data so that it is impossible to restore, even with un-deletion techniques. While the overwritten data will be irrevocably destroyed, Kessler [26] notes two weaknesses with this class of techniques. The first is that it may miss some data and, second, it may leave traces of wiping activity, most notably the presence of the wiping tool. A large portion of the existing Android anti-forensic literature is concentrated on artifact wiping.

Albano, et al. [3] describes a technique for sanitizing a phone by removing deleted files. Their technique works by booting a custom recovery image, copying all files to an external storage device, overwriting the entire internal memory and copying the files back. The recovery image is a minimal operating system image which was originally intended for performing a complete reset of the phone but which can be replaced on a rooted phone [21]. This procedure will make sure that any data previously deleted will now be irrevocably lost, but has the disadvantage of being an off-line procedure requiring significant time and manual effort.

Another pattern of design for artifact wiping, adopted by several researchers [7, 27], uses an application on an unmodified (or rooted) Android phone to detect the presence of a forensic analysis tool and start deleting data.

To trigger the wipe, two methods have been used which include reading the system logs [7, 27] and detecting a USB connection. Reading the system logs has the disadvantage of being slow, since it has to wait for the event to occur the log message to be generated, written and finally read back in before being able to take action.

Detecting a USB connection suffers from a lack of specificity, especially in earlier versions of Android. Version eight of the Android API (corresponding to Android 2.2 'Froyo' [5]) is unable to use the improved USB support introduced in API version 12 (Android 3.1 'Honeycomb MR1') and would need to be back-ported to API 10 (Android 2.3.4 'Gingerbread MR1') [6].

Regardless of the triggering mechanism being used, the anti-forensic application then has to delete data before the forensic analysis tool can extract it. All papers using this approach are concerned with this time window. They report measurements of time taken and how much can be overwritten in that time window prior to extraction.

## 2.3 Trail obfuscation

In the area of obfuscation, Albano, et al. [2] developed an automation system which can be used to make the phone act as if a user is present, thereby providing false evidence that the user was where the phone was at a particular time. They describe several systems based on generally available software automation, testing tools and finally built one of their own based on recording and playing back user interactions (e.g. touch screen input). For recording events, they connect the phone to a PC and perform the recording over a USB debug connection. Playback is performed either from the recording PC or on the phone from a script file uploaded from the PC. Tests show that this system can be used to post messages to Facebook and send SMS messages. After sending the messages, forensic analyses were performed, which showed no conclusive traces of the automation system on the phone. For their system to work, the phone had to be rooted, and for running without being connected to a PC, a general purpose scheduling application had to be installed. The researchers hint at, but do not state outright, that the uploaded script is sufficiently obtuse not to be a significant trace. The controlling PC was run entirely in RAM from a Linux live CD, and thus left no traces whatsoever.

## 2.4 Attacks against processes and tools

Procedures used by computer forensic analysts are supposed to follow the public guidelines set by central bodies (e.g., ACPO and NIST). It is, therefore, relevant to design anti-forensic tools to attack these procedures. An example would be hiding information where only a single, password protected program can access it, as in Distefano, et al.'s [12] design. In this case, the analyst can choose between following procedure and not getting the data or rooting the phone and getting the data and violating ACPO principle one, e.g., not modifying the data on the device.

Smartphones are complex, integrated devices, often necessitating the use of the entire original system in the analysis. This stands in contrast to personal computers, which consist of discrete components connected through standard interfaces which can be examined one by one, thereby bypassing some protection. For this reason, the published standards condone much more invasive examinations for mobile phones than for PCs [24]. However, many of these investigations take time and effort, are specific to individual phone models, require individual testing and require potentially in-depth explanations in court environments.

## 3. Methodology

This research aims to investigate the potential effectiveness of Android operating system-level anti-forensics modifications. The idea focuses on the modification of the operating system to deceive automated tools. Any tool could have been chosen for the examination of the hypothesis. As a matter of convenience, Cellebrite (version: App: 1.1.9.4 UFED, Full: 1.0.2.7, Tiny: 1.0.2.1) [8] and XRY (version: 6.1.1) [28] were used in this particular experiment. The phone that was used in the research was a HTC Desire running the CyanogenMod distribution of Android. The following steps were executed in this experiment.

The **first step** was to investigate operating system modifications. In order to achieve this, the source code for the CyanogenMod 7.2 community distribution of Android was downloaded, built and installed according to the CyanogenMod project's instructions [10]. Once the phone was running this version of CyanogenMod, modifications were introduced to trace the behavior of the forensics tools. Content providers are applications that wrap databases on the phone, performing security checks and format conversions as required by the Android specifications. On the assumption that both Cellebrite and XRY used content providers to access data on the phone, these modifications took the form of altering the content providers to generate logs of how they are called. The merits of this assumption are discussed further in section 4.1

The **second step** consulted Android documents for information on the returned data format for each call. According to the documentation for Cellebrite and XRY, they both used a USB debugging connection to extract data from Android phones. To determine what information is available for triggering anti-forensic behavior, a separate application was developed that produces event logs as noted by the operating system when this connection is established and severed. While a USB device (such as a phone) must identify itself to its host (i.e. a forensic analysis system), the device does not learn anything about the host [29]. This means that the phone cannot distinguish the connection of a forensic analysis system from the connection of a PC. However, Cellebrite and XRY (and according to Azadegan, et al. [7], Paraben Device Seizure [30] and Susteen Secure View [22]) work by uploading an application to the phone over Android's standard USB debug interface [11]. In addition, they also indicate that these applications export the data over the USB debug interface.

The **third step** modified the content providers to recognize when they were being called by forensics tools. This takes into account the information from the separate programs on how to recognize USB debugging and behavior specific to the tools. The idea is to make the content providers exhibit anti-forensic behavior when they detect the presence of the forensics tools, but still be sufficiently close to the original behavior for the forensics tools to believe in the data they receive.

The anti-forensic behaviors will inject a delay before returning any data, returning no data, returning data hardcoded in the content provider and/or returning data from an alternate database. The full range of anti-forensics will be implemented for the phone's contact list. To prove that the techniques can be generalized, modifications will be introduced to return no data for queries for the SIM contact list and SMS messages. When anti-forensics is used, the real data should not be extracted, the intended false data should be extracted and no errors should be reported.

The package manager will be modified to detect attempts by the forensics tools to install their applications on the phone, and reject the installation. In this case, the tools will be permitted to report errors to the forensic analyst, but no data should be returned.

The **fourth step** created and entered a dataset into the phone. The phone contacts were entered using the standard on-board tools, and contained two entries, each with a name and a phone number. The contacts were not synchronized with any other service.

The **fifth step** performed forensic extractions using both Cellebrite and XRY. The first extractions were conducted with the phone running an unmodified CyanogenMod operating system and then with each anti-forensic modification in turn. The results of the extractions were inspected for the real data, the false data for that anti-forensic case and any signs of the tools suspecting that something was wrong.

It should be noted that this research is intended to be a proof of concept that instigates a focused examination of contact artifacts using both logical and manual extractions. All other artifacts and interactions with the mobile phones are considered out of scope for this research.

# 4. Implementation and Results

The experimental work for this research consists of the three stages of examination, implementation and testing.

## 4.1 Examination

The content provider interface is the only way for forensics tools to gain access to information such as the contact list on an un-rooted Android phone. On a rooted phone, it would be possible for a forensics tool to bypass the content provider and read directly from the database, but this would require the tool to first find and interpret the database.

Under the assumption that the content provider is always used, instrumentation code was developed and inserted into it to write information about its behavior to the system log. For each call to the content provider's main query function, the code output the name of the calling program, the query arguments and which part of the existing program logic handled the query.

Both of the tools used the content provider in all observed cases. Cellebrite starts by making several queries to the raw contacts and settings modules, collecting general information such as the number of contacts and whether contacts are marked as deleted. It then goes through the raw contacts module, querying for information on each contact. For each contact, numerous queries are made for different kinds of information associated with it (name, phone number, email address, etc.). The extracted contact list matched the information entered and seen in the phone's built-in contact list application.

The program revealed that XRY makes relatively few queries in total, retrieving an entire data module with each query. The modules retrieved are raw contacts and data. It is possible for XRY to interpret this data using information found in the Android API reference manual [4]. The extracted contact list, seen in Figure 1 – XRY unmodified phone extraction, matched that entered and displayed in the phone's built-in contact list application.

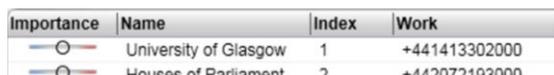

**Figure 1 -- XRY unmodified phone extraction**

## 4.2 Implementation

Previous work has triggered anti-forensic behavior either on finding log-entries relating to the installation of forensics tools or on connection of a USB cable. Reading logs requires waiting for the log messages to show up along with spending processing effort to read and interpret the log files. This is complicated by the fact that USB connections are frequent in everyday use which, potentially, generates a substantial amount of data to process.

One attribute common to the forensics tools used in this research is that they require the phone to be set to USB debug mode. This lets the tool control the phone and installed applications. Setting the phone to debug mode requires going deep into the settings menu and acknowledging a warning that 'USB debugging is intended for development purposes only'. It also removes the possibility to use the phone as USB memory as well as transfer files to and from it using standard file management tools. The fact that activating debug mode removes these normal and desirable features suggests that most users will not have debug mode activated. If so, triggering anti-forensic behavior upon activation of USB debugging will have a lower false-positive rate than triggering it on all USB connections. A stand-alone application was built to be a receiver for the intent broadcast when the USB state of the phone is changed. Using the extra information contained in the intent, the program determines whether the USB cable is connected or not, and whether the phone is in debugging mode.

Finding no data in a well-used phone would be suspicious. Arranging for the analyst to find plausible, but non-incriminating, data increases the chance of the analyst accepting the output and concluding that the phone holds no relevant data. While examining the behavior of the forensics tools and the operating system, sufficient data was obtained to determine the format for the extracted contact list. Using that information, false data in the correct formats was constructed and inserted into the code for the contacts provider. This data is to be returned in response to queries from each tool. As a fallback, if a query comes from an unknown tool and USB debugging is on, no results will be returned.

**4.3 Results**

The operating system was rebuilt to contain the module, uploaded to the phone and installed using the recovery image. This replaced the operating system already on the phone, but preserved user data such as contact lists and SMS messages. Each modified version of CyanogenMod was approximately 90 MB in size and took slightly less than one minute to install.

The phone was reinstalled using a standard CyanogenMod 7.2 operating system. The monitoring application will be installed and started. With the monitoring application running, the USB cable will be plugged and unplugged with the phone in both USB mass storage and USB debugging modes. The messages printed by the application should match the state of the USB cable and settings. Two screenshots from the running application are in Figure 2 - USB debugging on and in Figure 3 - USB debugging off.

Both images contain mostly raw information, so the processed information indicating USB connection and debugger state has been indicated with outlines. The debugger is indicated by its name "ADB", which stands for "Android Debug Bridge" [4] .

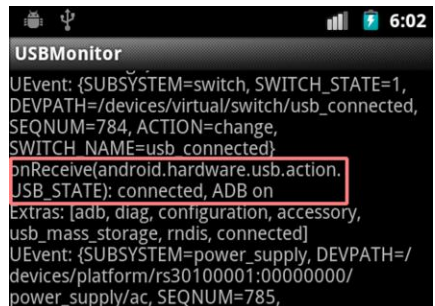
**Figure 2 - USB debugging on**

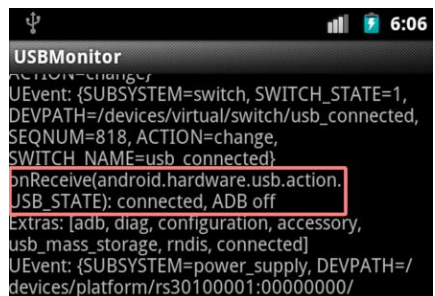
**Figure 3 - USB debugging off**

**4.3.1 Response delays.** The phone was reinstalled using a modified CyanogenMod 7.2 operating system. The only modification from default was the insertion of a delay into each query of the contact list provider. Extractions of the contact list were performed using Cellebrite and XRY and the delay increased until the tools presented errors instead of performing successful extractions.

Cellebrite has a low tolerance for response delays. It accepts a delay of five seconds at each call to query(), but ten seconds is enough to make it abort the extraction and show the error message in Figure 4 – Cellebrite error message.

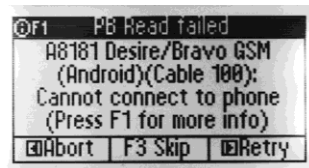
**Figure 4 - Cellebrite error message**

When this error happens, no report is created and no data is extracted, even if other information on the phone could have been extracted successfully. Adding a five second delay to each call lengthens the extraction time from approximately 25 seconds for an unmodified phone (Figure 5 – No delays) to approximately 2 minutes 20 seconds (Figure 6 – Delay 5 seconds per query call).

| Extraction start date/time: | 10/08/12 14:46:09 |
|---|---|
| Extraction end date/time: | 10/08/12 14:46:33 |

**Figure 5 - No delays**

| Extraction start date/time: | 10/08/12 14:10:38 |
|---|---|
| Extraction end date/time: | 10/08/12 14:13:00 |

**Figure 6 - Delay 5 seconds per query call**

**4.3.2 Rejecting installation of forensics tools.** The phone was reinstalled using a modified CyanogenMod 7.2 operating system. The only difference from the standard code was the modifications to the package manager for rejecting installation of applications named 'com.client.appA' or 'example.helloandroid' which previous experiments determined were the names used by the applications uploaded to the phone by Cellebrite and XRY, respectively. Extractions of the contact list were performed using Cellebrite and XRY. The extraction results were compared to results from an unmodified CyanogenMod.

Cellebrite completely failed to perform the extraction, instead presenting the error message in Figure 7 - Cellebrite installation rejected.

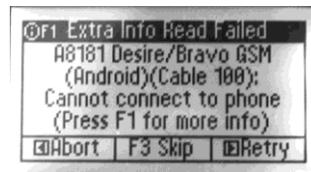

**Figure 7 - Cellebrite installation rejected**

XRY completed the extraction. However, a message indicates that an error had occurred and that the extraction was incomplete. The error log is displayed in Figure 8 – XRY extraction log. The log was not particularly clear on what went wrong and the extent of the consequences. The report was missing the sections for Device/App Usage, Contacts and Web/Bookmarks.

**4.3.3 Hardcoded false contact list.** The phone was reinstalled using a standard CyanogenMod 7.2 operating system. Using the built-in contact list application, two contacts were entered into the phone contact list (see Figure 9 – Contact list). Extractions of the contact list were performed using Cellebrite and XRY, and both extractions successfully displayed the correct contact information for both contacts.

| 7 | MAIN | Success | 11:45:58 | Processing device [HTC Desire A8181] connected to DummyPort []... |
|---|---|---|---|---|
| 8 | MAIN | Success | 11:45:58 | Starting process of ANDROID (6.1.1) |
| 9 | ANDROID | Success | 11:45:58 | Connecting |
| 10 | ANDROID | Success | 11:46:08 | Connected |
| 11 | ANDROID | Unsuccessful | 11:46:08 | Receive packet failed |
| 12 | ANDROID | Success | 11:46:12 | Device is rooted. Application data will be extracted. |
| 13 | ANDROID | Success | 11:46:14 | Extracting email data. |
| 14 | ANDROID | Success | 11:46:18 | Extracting Google Talk data. |
| 15 | ANDROID | Success | 11:46:30 | Disconnecting |
| 16 | MAIN | Unsuccessful | 11:46:35 | ANDROID (6.1.1) completed with error |
| 17 | MAIN | Success | 11:46:35 | Starting process of DISKSTOR (6.1.1) |

**Figure 8 - XRY extraction log**

The contact list database was copied from the phone to a PC. The phone was reinstalled using a modified CyanogenMod 7.2, the only difference from the standard operating system being the contact list provider, which contained hardcoded false data. Three sets of false data were tested.

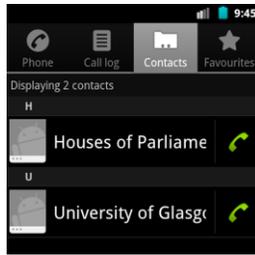
**Figure 9 - Contact list**

Cellebrite and XRY were provided with contact lists containing one contact each, this being the technical support phone number for each tool. Unknown tools will be provided with an empty contact list. Cellebrite and XRY were used, in turn, to extract the phone contact list. Both tools displayed the same number of contacts as in the real contact list, but each displayed the name and number to technical support for the respective tool. The display of false data is demonstrated in Figure 10 - Cellebrite extraction report and in Figure 11 - XRY extraction report.

**Phone Contacts**

Total Entries: 2
PBB MD5 Hash: A5F38C59FC9096D899352CEED848AF0F
PBB SHA256 Hash: 63BFC342 9078845 71A1907 2490120
D9CA979 83AB5D2 2BB3812 6A97718 8D08CCA

| #1 | Cellebrite Technical Support (Memory: Phone) |
|---|---|
| Work: | +495251546490 |

| #2 | Cellebrite Technical Support (Memory: Phone) |
|---|---|
| Work: | +495251546490 |

**Figure 10 - Cellebrite extraction report**

| Importance | Name | Index | Work |
|---|---|---|---|
| —○— | XRY Technical Support | 1 | +4687390270 |
| —○— | XRY Technical Support | 2 | +4687390270 |

**Figure 11 - XRY extraction report**

Finally to simulate the connection of an unknown forensics tool, the phone was connected to a PC still in USB debugging mode and the contact list inspected using the built-in application. In this case, the built-in application shows an empty contact list, as seen in Figure 12 - Empty contact list.

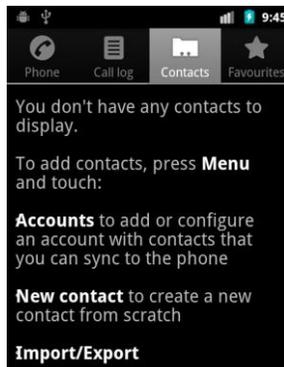
**Figure 12 - Empty contact list**

**4.3.4 Alternative Databases.** The phone was reinstalled using a modified CyanogenMod 7.2 operating system. The contact list provider was modified to return results from alternate databases depending on whether the query comes from Cellebrite, XRY or the phone itself outside of USB debugging mode.

When the contact list was entered manually, the phone software created an SQLite database file on the phone and stored it in /data/data/com.android. providers.contacts/databases/contacts2.db. This file was copied to a PC in two instances and changed to contain the technical support phone numbers for Cellebrite and XRY, respectively.

These changed database files were uploaded to the phone to be used by the modified contact list provider. These databases were stored in /data/data/com.android.providers. contacts/databases/cellebrite.db and /data/data/com.android.providers.contacts/databases/xry.db.

The extractions of the phones contact list were performed using Cellebrite and XRY. Neither tool saw the two real contacts. Both tools displayed a single number which is their own technical support phone numbers. The results are available in Figure 13 - XRY alternate database and in Figure 14 - Cellebrite alternate database.

| Importance | Name | Index | Work | Attachments |
|---|---|---|---|---|
| | XRY Technical Support | 1 | +4687390270 | 0 |

**Figure 13 - XRY alternate database**

**Phone Contacts**

Total Entries: 1
PBB MD5 Hash: 324B323C2612C3A2C685F72FB6C5641D
PBB SHA256 Hash: 34815D95 B25C782 B6387DF 426C414 DF75C2F 657167A 68BD636 DD1C085 8359767

| #1 | Cellebrite Technical Support (Memory: Phone) |
|---|---|
| Work: | +495251546490 |

**Figure 14 - Cellebrite alternate database**

**4.3.5 Delayed restoration.** The phone was reinstalled using a modified CyanogenMod 7.2 operating system. The contact list provider was modified to return results from alternate databases, as in the previous test. In addition, the provider was also modified to continue returning the same false contact list for thirty seconds after the USB cable had been unplugged regardless of which tool was used for the extraction. The purpose of this delay is to address a scenario where the analyst performs a manual check after the tool has finished its automatic extraction.

The experiment was re-run and displayed the same results from the previous test. For both tools, the phone was disconnected from the USB cable and the contact list inspected using the built-in contact list application. This inspection displayed the number for the respective technical support. After thirty seconds were allowed to pass, the built-in contact list application was re-opened. At that point, it showed the two contacts in the real contact list.

## 5. Conclusions

Conducting mobile phone forensics investigations is an increasingly challenging and complex undertaking. The interest in mobile device anti-forensics is increasing from within both academia and industrial environments. The complexity of the environment coupled with anti-forensics operating system modifications potentially inhibits mobile phone forensics investigation.

The results from this research demonstrate that it is possible to modify an Android operating system to present false information to the forensics tools. The forensics tools trust the phone software to return the correct results. The means that the tools are trusting file system drivers, the installation package manger and lower level functions.

The tools in this experiment performed logical acquisitions of the devices. Assuming the phone is supported, the tools in the experiment can be used to perform a physical acquisition of the phone's entire memory, thereby bypassing the high-level phone software and only trusting the phone's file system driver to return the correct files. However, using that mode of acquisition requires that the tool or the analyst perform data interpretation themselves, without the help of the phone software. Both tools also use standard methods of installing software, thereby trusting the package manager to install that software correctly. These high-level software packages, in turn, trust the lower levels to function correctly. Therefore, the forensics tools also trust, by extension, all lower levels of the Android stack, including the Dalvik virtual machine, the Linux kernel and the hardware. Any component of a system under forensic analysis that is trusted by one party is a point of attack for their opponent. Since the content providers and package manager are trusted components, they are natural targets for anti-forensics activities.

The research demonstrates that it is possible to modify components to present false data. This task is aided by the fact that Android is an open system, with specifications and source code that are freely available. Several projects use that source code to build community distributions of Android which can be installed on many different models of phones. The installation requires that the phone be rooted, which is possible to do on many phones and popular among technologically sophisticated consumers. Step-by-step guides, available on the Internet, describe how to root phones and install community distributions of Android. These community distributions depend upon contributions of code from the general public. They, therefore, make it easy to modify their code and install the modified versions. While programming skills are a prerequisite, it is possible to modify and replace content providers and the package manager. In this research, the OS components that were modified allowed for activity on the phone to be monitored and responses to be customized based on the state of the device.

In broad security terms, behavior that is repeated provides a foundation for identification. If that behavior is not the same as that produced by regular use, it presents an opportunity for anomaly detection. The tools in this experiment provided both. Every time the tool behavior was observed, each utilized the same specific name for the uploaded application. The tools also queried the content providers in the same way. They both also require the phone to be in USB debugging mode, which is unlikely to be the case for a phone in regular use.

The experiments performed here have shown that it is possible to distinguish between normal use and forensic analysis by looking at whether USB debugging is enabled, and that it is possible to distinguish between different forensics tools by looking at the names of their applications.

Mobile phone anti-forensics is concerned, to a large degree, with overwriting or deleting information. While this potentially makes the data completely unavailable to the forensic analyst, it also makes it unavailable to the legitimate user if the phone is eventually returned. Providing false data to the analyst presents the possibility of hiding the real data and reverting to it after the completed analysis.

The anti-forensics solutions implemented for this research demonstrates that it is possible, from a proof of concept perspective, to deceive a device implementing a logical acquisition. The real information is left in its place, while false information is fed to forensics tools from other sources. Forensics tools place a great deal of trust in the Android software, but that software can easily be modified and replaced. When suitably modified and replaced, that software can feed false information to the tool. Neither tool used in this research detects this subterfuge and presents the false information to the analyst as if it was real. The anti-forensics software modules are present on the phone and can be seen by the analyst should they do a logical extraction of the phone's file system. However, their presence and function is not obvious and, even if they are detected, reverse-engineering them potentially requires significant time and effort from the analyst. The results of this research empirically support the idea that examiners should not rely on a single tool or extraction method in their analysis of mobile devices.

## 6. Future Work

Future mobile phone anti-forensics work will examine additional options for triggering anti-forensics behavior at both the operating system and the hardware levels. Additional work should also investigate solutions at these levels that would counter additional extraction techniques like physical extraction or chip removal. More sophisticated triggers could include investigating single or multiple calls to areas of the phone such as raw_contacts. These calls could also be coupled with a series of other activities and/or states of the device. Monitoring queries on a mobile device in conjunction with the state of the device could provide insight into device data traffic patterns.

Presenting false data whenever the network connection is lost may be a valid anti-forensic strategy. The connection may be legitimately lost during everyday use, for example, by the user walking into an underground rail station. In this case, the phone functionality is unavailable to the user anyway, so the unavailability of phone-related data may not be a significant drawback.

The combination of destructive data techniques with operating system modification should be explored as well. The improved hiding and triggering properties found by implementing anti-forensics in the operating system over using a standalone application would also be able to hide destructive anti-forensics routines. For example, the package manager could be extended to not only reject the installation of forensics tools, but use the installation attempt as a trigger to perform a complete wipe of the phone. This would free the anti-forensics routines from the timing constraints which apply when they run as a separate application.

After various anti-forensics options have been explored, investigations will take place into how to effectively detect operating system modifications and the most efficient way for law enforcement to confront these issues. This includes investigating additional activities to trigger similar behaviors that are not necessarily intentional anti-forensics techniques. Additional work by the authors will examine the creation of custom ROMs that will introduce additional forensics capabilities along with exploring crossover issues with Bring Your Own Device (BYOD) solutions in organizations.